\documentclass[12pt]{iopart}
\usepackage{epsfig}
\usepackage{graphicx}
\usepackage{lscape}
\usepackage{longtable}

\begin{document}

\title[Extrasolar Rocky Planet Lithosphere?]{An aluminum/calcium-rich, iron-poor, white dwarf star: evidence for an extrasolar planetary lithosphere?}

\author{B. Zuckerman$^!$, D. Koester$^2$, P. Dufour$^3$\footnote{CRAQ Postdoctoral Fellow}, Carl Melis$^4$\footnote{Joint CASS Departmental and NSF AAPF Fellow}, B. Klein$^1$, and M. Jura$^1$}

\address{$^1$Department of Physics and Astronomy, University of California, Los Angeles, CA 90095, USA}
\address{$^2$Institut fur Theoretische Physik und Astrophysik, University of Kiel, 24098 Kiel, Germany} 
\address{$^3$D\'epartement de Physique, Universit\'e de Montr\'eal,Montr\'eal, QC H3C 3J7, Canada}
\address{$^4$Center for Astrophysics and Space Sciences, University of California, San Diego, CA 92093-0424, USA}

\begin{abstract}
The presence of elements heavier than helium in white dwarf atmospheres is often a signpost for the existence of rocky objects that currently or previously orbited these stars.  We have measured the abundances of various elements in the hydrogen-atmosphere white dwarfs G149-28 and NLTT 43806.  In comparison with other white dwarfs with atmospheres polluted by heavy elements, NLTT 43806 is substantially enriched in aluminum but relatively poor in iron.  We compare the relative abundances of Al and eight other heavy elements seen in NLTT 43806 with the elemental composition of bulk Earth, with simulated extrasolar rocky planets, with solar system meteorites, with the atmospheric compositions of other polluted white dwarfs, and with the outer layers of the Moon and Earth.  Best agreement is found with a model that involves accretion of a mixture of terrestrial crust and upper mantle material onto NLTT 43806.  The implication is that NLTT 43806 is orbited by a differentiated rocky planet, perhaps quite similar to Earth, that has suffered a collision that stripped away some of its outer layers.

\end{abstract}
\maketitle

\section{INTRODUCTION}

The study of extrasolar planetary systems has provided a host of surprises that include the realization that the photospheres of many white dwarf stars are the graveyards for rocky bodies that once orbited these stars (e.g., Jura 2008; Farihi et al. 2009; and references therein).   The photospheres present a tableau on which the elemental compositions of these erstwhile asteroids or rocky planets are displayed.  Study at optical and ultraviolet wavelengths of externally polluted white dwarf photospheres is a field still in its adolescence.  Nonetheless, even at these early times, it is already possible to explore the abundance of aluminum (an element of only moderate cosmic abundance) in rocky extrasolar bodies.  Aluminum is the 11th most abundant element by number in the Sun and the 8th most abundant element in CI chondrites (Lodders 2003).  

We derive the abundances of a suite of elements in the heavily polluted white dwarfs G149-28 and NLTT 43806; the photosphere of the latter star is both aluminum-rich and iron-poor.  Since Al is, after oxygen and silicon, the third most abundant element in Earth's  crust, a plausible explanation of the high abundance of Al in the atmosphere of NLTT 43806 could be accretion of crustal material from a differentiated extrasolar rocky planet or massive asteroid.  We evaluate a variety of objects and evolutionary processes that might lead to an Al-rich Fe-poor configuration and conclude that, indeed, accretion of a mixture of crustal and upper mantle material from a differentiated planet, perhaps quite similar in size to Earth, provides a credible explanation of the mixture of nine heavy elements detected in the atmosphere of NLTT 43806. 

\section{OBSERVATIONS AND ANALYSIS OF STELLAR PROPERTIES}

We used the HIRES echelle spectrometer (Vogt et al. 1994) on the Keck I telescope at Mauna Kea Observatory in Hawaii to observe white dwarfs WD1257+278 (G149-28) and WD1653+385 (NLTT 43806).  An observing log is given in Table 1.  Both stars are of type DAZ -- showing the presence of hydrogen and of elements heavier than helium.  Our 2007 and 2010 observations covered the range 3130 to 5940 \AA.  The blue cross disperser was combined with a 1.15" slit resulting in a spectral resolution $\sim$40,000.  Reduction procedures utilized two software packages, IRAF and MAKEE.  

G149-28 was found to be a DAZ by Zuckerman et al. (2003) who also used the HIRES spectrometer and covered the range 3700 to 6700 \AA.  
We are aware of three determinations of effective temperature (T$_{eff}$) and gravity (log g in cgs units): 8491 K and 7.90  (Zuckerman et al. 2003), 8733 K and 8.328 (Holberg et al. 2008) and 8710 K and 8.36 (Limoges \& Bergeron 2010).  The first reference used  UBVRIJHK photometry, the other two are spectroscopic determinations with very similar models and methods.  We use the Sloan Digital Sky 
Survey (SDSS) spectrum for our own spectroscopic analysis.   Employing the Balmer lines, including H$\alpha$, we obtain T$_{eff}$ = 8596 
K, log g = 8.12.  As can be seen in Figure 1, H$\alpha$ is only poorly fitted.   Excluding H$\alpha$ gives a better fit and T$_{eff}$ = 8667 K, log 
g= 8.21. This is in reasonable agreement with the most recent determination by Limoges \& Bergeron, who use a blue spectrum without H$\alpha$.  Another method is to use the measured parallax ($\pi$ =  0.0289$\pm$0.0041 from Van Altena et al. 1994, as given in Holberg et al. 2008), together with UBVRI (Holberg et al. 2008) and JHK (Zuckerman et al. 2003) photometry.  Fitting these with theoretical magnitudes derived from our model grid, we obtain T$_{eff}$ = 8593$\pm$142 K, log g = 7.95$\pm$0.04. If the parallax is assumed to be correct, the parameters are obviously tightly constrained.   To estimate the influence of the uncertainty in parallax, we have made the same fit assuming $\pi$ increased and decreased by 1$\sigma$. The two solutions are 8577/8.14 and 8601/7.71. The best fit is obtained for the first of these pairs. Considering the various results, for G149-28 we adopt as final parameters T$_{eff}$ = 8600$\pm$100 K, log g = 8.10$\pm$0.15.

The status of NLTT 43806 requires some discussion.  The metal-polluted nature of this white dwarf was first investigated by Kawka \& Vennes (2006) who used a moderate resolution grating spectrometer.  They characterized the star as a DAZ (H$\alpha$ and H$\beta$ are seen in the spectrum) and derived an effective temperature of 5700 K and the following (logarithmic) elemental abundances by number: [Ca/H] = -8.4, [Mg/H] = -7.0, and [Na/H] = -8.1.   

Kilic et al. (2009) obtained near-infrared spectroscopy of NLTT 43806 (they refer to the star as J1654+3829).  From a fit to the broadband spectral energy distribution, they deduce that NLTT 43806 is a DZA with T$_{eff}$ = 5830 K and He/H = 6.2 by number.  We tried to match the HIRES measured line profiles to a star with the Kilic et al temperature and helium abundance, but were unable to do so.

In the absence of any positive evidence for the presence of significant quantities of He in the atmosphere of NLTT 43806, we treat the star as having an H-dominated atmosphere; Giammichele, Bergeron, \& Dufour (in preparation), find a very good fit to the H$\alpha$ line profile plus (SDSS and 2MASS) ugriz+JHK photometry for a log g = 8 (fixed), 5900 K pure hydrogen model.  Our high-resolution HIRES spectra demonstrate the weakly magnetic nature of NLTT 43806 (Figures 2-4).  We used an analysis similar to that described in Section 3.2 of Farihi et al (2011b) and found a good fit to the line profiles with a uniform B field of 70 kG (Section 3). This field is too weak to effect the determination of the atmospheric parameters derived by Giammichele et al.  NLTT 43806 thus joins a few other known weakly magnetic white dwarfs that are polluted with heavy elements: G77-50, G165-7, LHS 2534, and LTT 8381.

\section{ELEMENT ABUNDANCES}

For G149-28 and NLTT 43806 a sequence of models was calculated with the T$_{eff}$ and log g given in Section 2.  A DAZ model for G149-28 was calculated with these parameters and the abundances of the observed elements Ca, Mg, Al, Fe, Ti, and Ni were varied until the model lines had approximately the same strength as the observed ones.  This model was used as the starting point for the analysis of element abundances.  All observed element abundances (relative to H) were then varied between -0.6 dex and +0.6 dex of the previously determined starting abundances, in steps of 0.2 dex.  Theoretical equivalent widths (EW) for the observed lines were calculated from these models.  An element abundance was obtained for each individual line by matching the measured and model EW (Table 2).  Note that all [Z/H] given in Table 2 refer to the total abundance by number of a given element and not to the abundance of the listed neutral or ionic stage;  the element abundances come from a weighted average of the abundances derived from individual lines.   One source of error in the derived abundances obtains from the standard deviation of the distribution of abundances derived from the individual lines.  We also included in the error budget the change in abundances that result from changing T$_{eff}$ by 100 K and log g by 0.15, our adopted uncertainties in these parameters (Section 2).  The various potential sources of error were added in quadrature. 

For NLTT 43806, for each element, models including a magnetic field were computed for log [Z/H] between -7 and -10.5 in steps of 0.5 dex.  Model grids including two elements were used for blended lines (e.g., Fe+Mg, Ca+Al, Ti+Fe).  The fitting method is described in Farihi et al (2011b) and Melis et al (2011).  For simplicity, a constant B field was assumed in the model spectrum calculations; a field of 70 kG gave generally good agreement between the model and the data (Figures 2-4).  The relative strengths of the $\pi$ and $\sigma$ ($\delta$m = 0, $\pm$1, respectively) Zeeman components would likely be somewhat different for a more realistic dipole field geometry, but the fits are sufficiently good that  this should have little effect on relative element abundances.  Also, the average of two different epochs (Table 1) could somewhat wash out the spectral effects of a non-uniform B field.  Table 3 lists total element abundances by number of atoms derived with our models based on the listed (magnetically split) transitions in NLTT 43806 and Figures 2-4 display the spectra and model fits to some representative lines.  

For each transition in NLTT 43806 we produced synthetic spectra with abundances that differed by $\pm$0.15 dex from the best fitting model; for many lines it is obvious that the model sensitivity to element abundance is substantially better than 0.15 dex.  Two of the authors independently estimated the abundances of the 9 detected elements giving their own weightings to the various model-fit transitions.  For each element the two estimates of abundance agreed precisely, with the exception of [Ti/H] that differed by (only) 0.05 dex.  We thus estimate an uncertainty in the abundance of each element of about 0.1 dex based on the dispersion in the abundances deduced from the model fit to various lines of a given element, except for Cr for which we use 0.15 dex.  As in the G149-28 error budget, we also included abundance changes that would result if the stellar parameters should differ from those we use in our preferred model (i.e., T$_{eff}$ = 5900 K; log g = 8.0).  Specifically we calculated models with (1) T$_{eff}$ = 6020 and 5750 K (these correspond to uncertainties in T$_{eff}$ from photometry) and log g = 8, and (2) T$_{eff}$= 5900 K and log g = 8.15.  The abundance uncertainties listed in the [Z/H] column of Table 3 are then the uncertainties from the model fit and stellar parameters added in quadrature.  

As can be seen from the size of the error bars on [Z/H], the contribution from a change in the stellar parameters is typically as large or larger than the 0.1 dex from the model fit uncertainties.  However, a change in a stellar parameter (T$_{eff}$ or log g) always is accompanied by changes of all element abundances (relative to H) in the same direction (e.g., increasing T$_{eff}$ increases all Z/H values and vice-versa for a decrease in T$_{eff}$).  Thus, in Tables 5-7 the listed element abundance ratios that do not involve H have uncertainties that are smaller than would be deduced from a simple combination of uncertainties on [Z/H] listed in Table 3.  For example, the ratio of the abundance of any two elements (where neither is H) changes by only 0.06 dex when log g changes by 0.15.  This 0.06 dex may be compared with the level of agreement in Table 6 between the Earth lithosphere model (Section 4.2) and the NLTT 43806 abundances which is often no better than 0.3 dex.  Thus, to significantly degrade the quality of agreement (or disagreement) between the lithosphere model and the NLTT 43806 abundance ratios, log g would have to be very much different from 8.0.   Various element abundance ratios in the other models we consider in Tables 6 and 7 and in Section 4 disagree with the measured NLTT 43806 abundance ratios by even more than 0.3 dex; thus plausible uncertainties in log g do not quantitatively impact the discussion (Section 4).

\section{DISCUSSION}

\subsection{Current Mass Accretion Rates}

To relate photospheric element abundances to relative abundances in a
rocky parent body or bodies requires knowledge of whether a given polluted
white dwarf is in the "building-up", "steady-state", or "declining" phase
of accretion.  These terms are most clearly defined and visualized in a
situation where only a single parent body is or has been accreted onto a
star.  During the building-up phase, matter is being accreted onto the star
but insufficient time has passed for any significant amount to have
settled out of the photosphere (which typically is contained within a
surface convective zone).  In a steady-state, the rate of accretion
is balanced by the rate at which material diffuses out of the convection zone.  
During the declining phase, accretion effectively has ended and remaining heavy
elements diffuse out of the photosphere. These three
phases are considered in some detail by Koester (2009), Jura et al. (2009),
and Zuckerman et al. (2010).

The ratio of settling times for G149-28 and NLTT 43806 (Tables 2 and 3) to their cooling ages are of order 5 x 10$^{-7}$ and 4 x 10$^{-6}$, respectively.  Thus, to catch these heavily polluted stars during the building-up phase is very unlikely.  While the declining phase can last a few settling times, this is still a very short period with respect to the cooling ages.  In addition, in the declining phase heavier elements such as Ni and Fe should be underabundant with respect to lighter elements because the former diffuse out of the convection zone faster than the latter.  In G149-28 the ratio of abundances of heavy to light elements is similar to CI meteorites and bulk Earth (Tables 2, 5, and 6), but the situation in NLTT 43806 is more complicated and is discussed in Section 4.3.

Mass accretion rates are discussed in some detail in Koester (2009), Jura et al (2009), Klein et al (2010) and Zuckerman et al (2010).  As noted by Klein et al., for rocky objects in the solar system most mass is carried by oxides of Si, Mg and Fe, and possibly also by metallic Fe and/or water.  However, based on optical spectra we are unable to measure the abundance of O in white dwarfs as cool as G149-28 and NLTT 43806.  Therefore, following Klein et al (2010), we assume that oxygen is carried primarily in oxides of abundant elements and that the rate of accretion of oxygen is equal to 1/2 the sum of the accretion rates (dM/dt) of all the elements listed in Tables 2 and 3.  Assuming that accretion is in the steady-state phase, we derive total mass accretion rates of 3 x 10$^8$ and 6 x 10$^8$ g s$^{-1}$  for G149-28 and NLTT 43806, respectively.  In comparison, utilizing a different procedure, Farihi et al (2009) derived rates of 3 x 10$^8$ and 10$^9$ g s$^{-1}$ for these two stars.  These rates could be underestimates if either or both of these stars have accreted a large quantity of oxygen bound in the form of H$_2$O.  As may be seen in Table 4 and Figure 14 of Farihi et al, these are among the largest accretion rates known for any white dwarf stars with a temperature less than 10,000 K; indeed NLTT 43806 has the largest known accretion rate in this temperature range.  One aspect of its large accretion rate is manifested by the stars listed in our Table 4.  This table illustrates that optical detection of numerous elements is most often feasible in warm (T $>$10,000 K) white dwarfs with atmospheres whose mass is dominated by helium; NLTT 43806 is the only exception to this pattern.   

\vskip 0.2in

\subsection{A Differentiated Rocky Planet in Orbit Around NLTT 43806?}

In Tables 5 and 6 we compare the element abundances in the atmospheres of G149-28 and NLTT 43806 with abundances in various other heavily polluted white dwarfs and solar system objects. The accreted material in G149-28 may have originated in an object with composition similar to bulk Earth or CI meteorites.  By contrast, the element abundances in NLTT 43806 are quite unusual, being marked by a large amount of Al and relatively little Fe.  Aluminum is abundant in various solar system environments including Earth's crust, Eucrite and Howardite meteorites, and refractory (Al- and Ca-rich) inclusions in chondrites.  Generally, Al can be quite abundant in rocky objects formed at high temperatures, whether in our planetary system or others.

Bond et al. (2010) simulate the formation of terrestrial planets in extrasolar planetary systems.  Planets that form at high temperatures (small semi-major axis) are enriched in refractory elements such as Al and Ca that are abundant in NLTT 43806.  However, such planets are largely devoid of relatively volatile elements such as Na which, in the atmosphere of NLTT 43806, has an essentially normal (i.e., CI meteorite or bulk  Earth) abundance (Table 6).  Similar considerations apply to the Ca- and Al-rich inclusions (CAIs) in our solar system; these display Al, Mg and Na abundances distinctly different (e.g., MacPherson \& Davis, 1993; Grossman et al. 2008) from those seen in NLTT 43806.  Therefore, the high Al and Ca abundances in NLTT 43806 cannot be explained simply as due to formation of a rocky body at high temperatures.

As may be seen in Table 6, among meteorites, element abundances in the common CI class are quite different from those seen in NLTT 43806.  A match with the less common Eucrite and Howardite classes is much better, with the exception of Na which is substantially more abundant in NLTT 43806.  Below we consider the nature of a parent body for these latter two classes of meteorites, as might be relevant for the history of NLTT 43806.

After O and Si, Al is the most abundant element in Earth's crust.  However, as may be seen in Table 6, the mixture of elements in the atmosphere of NLTT 43806 and in Earth's crust differ substantially.  A far better match can be achieved by a mixture (by weight) of 30\% continental crust and 70\% upper mantle (see Table 6 and Figure 5).  In such a model, seven of the listed elemental abundance ratios for steady-state accretion onto NLTT 43806 agree within a factor of two, while [Ni/Al] is about 5 times larger in NLTT 43806, when compared with the abundance ratios in the 30/70 mixture.

We envision a model in which the material now accreting onto NLTT 43806 was initially
ejected into circumstellar orbits by a collision of a differentiated Earth-like planet with another rocky
object. On Earth the mass of continental plus oceanic crust is about 0.47\% of Earth's
mass, or 2.8 x10$^{25}$ g, while the mass of the upper mantle is 22 times the mass of the crust
(Anderson 1989; 2007). So as not to take off too much mantle along with the crust, we presume
that the collision was probably a glancing one and assume that it removed about 3\% of
the crust along with about twice as much material (by weight) from the upper mantle. If ultimately 10\% of this circumstellar debris were to find its way onto NLTT 43806, then the total accreted mass would be $\sim$3 x 10$^{23}$ g. At a rate of 6 x 10$^8$ g s$^{-1}$,
accretion could be sustained for a few times 10$^7$ yr. This is one percent of the cooling age of
NLTT 43806 (2.5 x 10$^9$ yr, Bergeron et al 1995) and a plausible time span for rapid accretion given that $\sim$5\% of the DAZ (4 of 82) in the Zuckerman et al (2003) sample have mass accretion rates $>$10$^8$ g s$^{-1}$. 

An alternative to stripping part of the lithosphere off an earth-like world would be to strip the 
basaltic crust of a large, differentiated, asteroid such as Vesta; such material could have the composition of meteorites of the Eucrite or Howardite classes.  The mass of Vesta is 2.7 x 10$^{23}$ g
and we assume that the outer basaltic layer has a mass of 10$^{22}$ g. If a collision ejects 10\%
of this "crust" into circumstellar orbits and 10\% of that debris, or 10$^{20}$ g, is ultimately
accreted onto NLTT 43806, then at 6 x 10$^8$ g s$^{-1}$, accretion could be sustained for only $\sim$5000
years. Given the cooling age of the white dwarf and the number of white dwarfs that
astronomers have investigated with reasonably high signal-to-noise spectra, detection of
such a short-lived event would be quite improbable.

Thus, accretion onto NLTT 43806 of collisional debris originating in the outer layers of a differentiated rocky planet is a model in agreement with the various constraints imposed by our observations.

\vskip 0.2in

\subsection{Extreme Mass Accretion Rates In a Declining Phase Model} 

As noted in Section 4.1, when a polluted white dwarf is in the declining phase, in the convection zone heavy elements such as Ni and Fe become less abundant relative to lighter elements such as Al.  As may be seen in Table 6, this is the situation in the photosphere of NLTT 43806.  The absence of evidence for a dusty debris disk is also consistent with a declining phase condition.  We therefore consider a model in which we assume that the parent body that polluted the white dwarf atmosphere had a ratio by number of Fe to Al of 10.5  -- the ratio in CI meteorites -- and we deduce for how long NLTT 43806 has been in the declining phase.
With the relative abundances and settling times given in Table 3, we find that 6.6 x 10$^4$ years ago [Fe/Al] in the atmosphere of NLTT 43806 would have been $\sim$10.  At that time, in the steady state phase, the mass accretion rate of Fe would have been 735 times larger than now and, for Al, $\sim$50 times larger than at present. Increases in accretion rates of the seven other detected elements compared to those listed in Table 3 would have ranged from 25 for Na to 1180 for Ni.

Table 7 lists what abundances in the atmosphere of NLTT 43806 would have been, relative to aluminum, 6.6 x 10$^4$ years ago.  While the declining phase picture fixes the problem of the low Fe abundance in NLTT 43806, it does not improve the situation for Mg/Al or Si/Al.   As may be seen, when comparison is made between NLTT 43806 and bulk Earth and CI meteorites, these important ratios remain as discrepant as before (see Table 6 and Figure 5).   Agreement with abundances in GD 362 appear to be somewhat better, but overall still not as good as those in the lithospheric model discussed in Section 4.2.

In addition, 6.6 x 10$^4$ years ago the total mass accretion rate would have been huge, $\sim$10$^{11}$ g s$^{-1}$.  This is an order of magnitude greater than the accretion rate for GD 362 (Jura et al. 2009) and comparable to that for SDSS J0738+1835 (Dufour et al. 2010), the two previously known champions.

Even with such a huge previous accretion rate, some detected elements such as Ti, Cr and Ni could not have been seen for very many settling times given how weak their lines appear in the HIRES spectra.  We estimate that if these elements were only 4 times less abundant in the photosphere than they are now, then they would not have been detected.  Thus, the available time for recovering a spectrum similar to that observed with HIRES would be only about 8 x 10$^4$ years.  This time span is only $\sim$3 x 10$^{-5}$ as long as the cooling age of NLTT 43806.   

Based on the above considerations, we cannot entirely rule out a declining phase model for NLTT 43806, but it would entail extreme conditions and low probability events.  A final definitive choice between the lithospheric model presented in Section 4.2 and the declining phase model could be supplied by detection or lack thereof of relatively heavy, yet relatively volatile, elements such as Mn and K that likely would be substantially more abundant in the former model.

\vskip 0.2in

\subsection{Comparison of NLTT 43806 With White Dwarf GD61}

Farihi et al (2011a) reported the presence of excess infrared emission implying the existence of a dusty debris disk in orbit around the highly polluted DBZA white dwarf GD61.  NLTT 43806 shows no evidence for orbiting dust grains (Farihi et al. 2009).   Other major differences between these two stars include the apparent dominant atmospheric constituent, H for NLTT 43806 and He for GD61, effective temperatures (5,900 K vs 17,300 K) and cooling ages (a few times 10$^9$ vs $\sim$10$^8$ yrs).  

Notwithstanding these major differences, the heavy element pollution of the two stars appears to be quite similar.  Specifically, the ratio of the four heavy elements detected in common -- Mg, Si, Ca, Fe -- is similar in the two stars, with Fe underabundant relative to Mg and Si.  In addition, the atmospheric residence time (before sinking) of heavy elements in the two stars is similar.  Because of the presence of the dust disk Farihi et al (2011a) consider that GD61 is in the steady-state phase and not the declining phase.  Therefore, Fe was underabundant in the portion of the object that supplied the observed atmospheric pollution, which they suggest may have been the outer layers of a differentiated parent body.  Their model for GD61 is thus similar to our lithospheric model for NLTT 43806.  However, in the case of GD61, Al has not yet been detected.  Because Al should be overabundant in the accreted debris at GD61 (if the Farihi et al model is appropriate), it is important to improve their weak upper limit to the Al abundance.  However, even if Al is found to be overabundant in GD61, as it is in NLTT 43806, without a measurement of the Na abundance in GD61, a viable alternative to the Farihi et al model would still exist; this is formation of a rocky asteroid or planet at high temperatures, as considered above in Section 4.2 and as discussed by Bond et al (2010), and accretion of that rocky object.

The total steady-state accretion rate required to sustain the pollution measured at GD61 is similar to that we estimated in Section 4.1 for NLTT 43806.  In Section 4.2 we showed that stripping off the outer layers of a differentiated massive asteroid probably yields too little mass to explain the pollution of NLTT 43806.  Given the rapid mass accretion onto GD61 it seems more likely that, also, for this star the outer layers of a differentiated planet are a more plausible source of material than are those of an asteroid.   However, because of the shorter cooling time and generally higher level of pollution of warm (young) white dwarfs compared to cool (old) ones (Zuckerman et al 2010), this argument in favor of a planet parent body and against a massive asteroid is not as strong for GD61 as it is for NLTT 43806.

\section{ CONCLUSIONS}

We have measured the ratio of element abundances in the heavily polluted DAZ white dwarfs G149-28 and NLTT 43806.  The latter star is weakly magnetic and is accreting material richly endowed with aluminum but poor in iron.    We compare the abundances of 9 elements in the atmosphere of NLTT 43806 with a wide variety of solar system and extrasolar objects (the latter both observed and postulated).  To calculate the relative abundances of these elements we assumed that log g = 8 and we adopted a simple model of a uniform B-field.  While it will certainly be desirable to measure the parallax and thus log g of NLTT 43806, our analysis implies that the abundance ratios of the 9 elements are insensitive to plausible variations in log g and in B.  

As may be seen in Klein et al (2011) and in Table 5 and references listed therein, the abundance ratios of the most abundant heavy elements in many externally polluted white dwarfs are not too dissimilar from these ratios in bulk Earth or the Sun.  In contrast, the Fe/Al ratio is occasionally very peculiar and can range widely; for example, Fe/Al is at least 100 times smaller in NLTT 43806 than it is in GALEX 1931.

To elucidate the nature of the parent body responsible for the pollution of NLTT 43806 the spirit of the analysis in the present paper has been to search for an object with similar element abundances -- either a known solar system object or portion thereof or a theoretical abundance model for plausible rocky objects in extrasolar planetary systems. Two such explanations for the abundance pattern seen in NLTT 43806 that we consider in the present paper are mentioned in the two paragraphs that follow.  But ultimately one must also entertain the possibility that the parent bodies in systems like NLTT 43806 and GALEX 1931 could have compositions dissimilar from any rocky object previously known or modeled.

One explanation for the unusual NLTT 43806 abundance ratios, especially its low iron abundance, would place the white dwarf in the "declining" phase of accretion.  In this phase accretion of orbiting material has ended and relatively heavy elements such as Fe diffuse out of the outer convective zone more quickly than do lighter elements such as Al.  However, this model would involve an extreme earlier accretion event and a short observable lifetime.
 
Noting that Al is the third most abundant element in Earth's crust, a plausible alternative to the declining phase model postulates that the 9 observed heavy  elements originated in the outer layers (lithosphere) of a differentiated rocky planet.  In such a model, the material was blasted off the planet by a collision with another rocky object. Such a collision would generate debris with a wide range of orbital eccentricities and semimajor axes, and could thus naturally explain how some of the aluminum-rich material could find its way
within the tidal radius of the white dwarf and eventually into its atmosphere.   A similar picture could account for heavy element pollution
of the white dwarf GD 40.  For this star Klein et al (2010) postulated a model of
accretion of a differentiated asteroid that lacked its outer layers -- just the inverse of the
NLTT 43806 situation where it is only the outer layers that have been accreted. In the
case of GD 40 a collision that could strip off the outer asteroidal layers may have played
the additional role of placing the remaining interior portion into an unstable orbit with
respect to a neighboring major planet -- thus facilitating an eventual orbital intersection
with the tidal radius of the white dwarf.

\vskip 0.2in

We thank Alan Rubin, John Wasson and An Yin for helpful discussions and the referee for helpful comments. This research
was supported by grants from NASA and NSF to UCLA. C. Melis acknowledges support from the NSF under award AST-1003318.
The data presented herein were obtained at the W.M. Keck Observatory, which is operated as a scientific partnership
among the California Institute of Technology, the University of California and the
National Aeronautics and Space Administration. The Observatory was made possible by
the generous financial support of the W.M. Keck Foundation.

\section*{References}
\begin{harvard}
\item[Allegre, C., Poirier, J.-P., Humler, E. \& Hofmann, A. 1995, Earth Planet. Sci. Lett. 134, 515]
\item[Anderson, D. 1989, "Theory of the Earth" (Boston, Blackwell Scientific Publications)]
\item[Anderson, D. 2007, "New Theory of the Earth", (Cambridge, Cambridge University Press)]
\item[Bergeron, P., Wesemael, F. \& Beauchamp, A. 1995, PASP 107, 1047] 
\item[Bond, J., O'Brien, D., Lauretta, D. 2010, ApJ 715, 1050]
\item[Dufour, P., Kilic, M., Fontaine, G., Bergeron, P., LaChappelle, F.-R., Kleinman, S. \& Leggett, S. 2010,]
\indent ApJ 719, 803
\item[Farihi, J., Brinkworth, C., GŠnsicke, B., Marsh, T.,  Girven, J., Hoard, D. W., Klein, B., \& Koester, D.]
\indent  2011a, ApJ 728, L8
\item[Farihi, J., Dufour, P., Napiwotzki, R. \& Koester, D. 2011b, MNRAS, in press]
\item[Farihi, J., Jura, M. \& Zuckerman, B. 2009, ApJ 694, 805]
\item[Grossman, L. et al. 2008, Geochim. Cosmochim. Acta 72, 3001]
\item[Hawkesworth, C. \& Kemp, I. 2006, Nature 443, 811] 
\item[Holberg, J., Bergeron, P. \& Gianninas, A. 2008, AJ 135, 1239]
\item[Jarosewich, E. 1990, Meteoritics 25, 323]
\item[Jura, M. 2008, AJ 135, 1785]
\item[Jura, M., Muno, M., Farihi, J. \& Zuckerman, B. 2009, AJ 699, 1473]
\item[Kawka, A. \& Vennes, S. 2006, ApJ 643, 402]
\item[Kilic, M., Kowalski, P. \& von Hippel, T. 2009, AJ 138, 102]
\item[Kitts, K. \& Lodders, K. 1998, Meteoritics \& Planetary Science 33, A197]
\item[Klein, B. 2011, Ph.D. Dissertation, UCLA]
\item[Klein, B., Jura, M., Koester, D. \& Zuckerman, B. 2011, ApJ submitted]
\item[Klein, B., Jura, M., Koester, D., Zuckerman, B. \& Melis, C. 2010, ApJ 709, 950]
\item[Koester, D. 2009, A\&A 498, 517]
\item[Limoges, M. \& Bergeron, P. 2010, ApJ 714, 1037]
\item[Lodders, K. 2003, ApJ 591, 1220]
\item[MacPherson, G. \& Davis, A. 1993, Geochim. Cosmochim. Acta 57, 231]
\item[Melis, C., Farihi, J., Dufour, P., Zuckerman, B., Burgasser, A., Bergeron, P., Bochanski, J. \& Simcoe, R.] 
\indent 2011, ApJ in press
\item[Vennes, S., Kawka, A. \& Nemeth, P. 2010, MNRAS 404, L40]
\item[Vogt S. et al. 1994, SPIE 2198, 362]
\item[Zuckerman, B., Koester, D. Melis, C., Hansen, B. \& Jura, M. 2007, ApJ 671, 872]
\item[Zuckerman, B., Koester, D., Reid, I.N., \& Hunsch, M. 2003, ApJ 596, 477]
\item[Zuckerman, B., Melis, C., Klein, B., Koester, D. \& Jura, M. 2010, ApJ 722, 725]

\end{harvard}

\clearpage
\begin{table}
\caption{Observation Log at the Keck Telescope}
\begin{tabular}{@{}lcccc}
\br
White Dwarf& Name& V mag& UT Date& Exposure (s)\\
\mr
WD1257+278& G149-28& 15.4& 2010 Mar 27& 1800 \& 1900\\
&  &  & 2010 Mar 28& 2400\\
WD1653+385& NLTT 43806& 15.9& 2007 May 6& 3000\\
&   &  &  2010 Jul 5& 2 x 2400\\
\br
\end{tabular}
\end{table}

\clearpage

\begin{table}
\caption{G149-28 Absorption Lines, Element Abundances, and Mass Accretion Rates}
\begin{tabular}{@{}lccccc}
\br
Element& Wavelength& EW& log [Z/H]& log(t$_{set}$)& log(dM/dt) \\
& (\AA) & (m\AA) & & (yr)& (g s$^{-1}$) \\
\mr
Mg I& 3833.4& 23$\pm$3& -7.46$\pm$0.05& &\\
Mg I& 3839.4& 58$\pm$5& -7.16$\pm$0.04& &\\
Mg I& 5174.1& 24$\pm$3& -7.15$\pm$0.05& &\\
Mg I& 5185.0& 38$\pm$6& -7.18$\pm$0.08& &\\
Mg & & & -7.24$\pm$0.15& 2.76& 7.67\\
Al I& 3945.1& 11$\pm$2& -8.16$\pm$0.08& &\\
Al I& 3962.6& 13$\pm$2& -8.17$\pm$0.06& &\\
Al & & & -8.17$\pm$0.09& 2.74& 6.81\\
Ca II& 3159.8& 35$\pm$10& -8.14$\pm$0.17& &\\
Ca II& 3180.3& 39$\pm$10& -8.33$\pm$0.15& &\\
Ca II& 3934.8& 460$\pm$20& -7.99$\pm$0.04& &\\
Ca I& 4227.9& 45$\pm$5& -8.11$\pm$0.06& &\\
Ca & & & -8.04$\pm$0.16& 2.77& 7.08\\
Ti II& 3350.4& 36$\pm$4& -9.34$\pm$0.06& &\\
Ti II& 3362.2& 12$\pm$2& -9.68$\pm$0.08& &\\
Ti II& 3373.8& 10$\pm$3& -9.62$\pm$0.15& &\\
Ti & & & -9.48$\pm$0.17& 2.64& 5.85\\
Fe I& 3571.1& 24$\pm$4& -7.52$\pm$0.09& &\\
Fe I& 3582.2& 39$\pm$2& -7.34$\pm$0.04& &\\
Fe I& 3609.9& 15$\pm$5& -7.40$\pm$0.17& &\\
Fe I& 3721.0& 23$\pm$2& -7.48$\pm$0.05& &\\
Fe I& 3735.9& 34$\pm$2& -7.43$\pm$0.04& &\\
Fe I& 3746.6& 14$\pm$2& -7.47$\pm$0.07& &\\
Fe I& 3750.6& 34$\pm$3& -7.37$\pm$0.06& &\\
Fe I& 3759.3& 8$\pm$2& -7.75$\pm$0.12& &\\
Fe I& 3816.9& 16$\pm$5& -7.35$\pm$0.16& &\\
Fe I& 3821.5& 29$\pm$2& -7.35$\pm$0.04& &\\
Fe I& 3827.0& 15$\pm$2& -7.47$\pm$0.07& &\\
Fe I& 3861.0& 20$\pm$5& -7.32$\pm$0.14& &\\
Fe & & & -7.41$\pm$0.15& 2.65& 7.97\\
Ni I& 3415.7& 19$\pm$3& -8.33$\pm$0.08& &\\
Ni & & & -8.33$\pm$0.11& 2.60& 7.13\\
\br
\end{tabular}
\end{table}
\noindent Notes $-$ Wavelengths are in vacuum.  Abundance ratios are by number of atoms.  The derived total abundance of each element appears in the entries for which the ionization state of an element is not indicated.  The settling times (t$_{set}$) are the diffusion times (t$_{diff}$) as defined in Koester (2009).  The mass accretion rates, dM/dt,  are applicable in a steady state situation (see Section 4).  Upper limits for [Z/H] for six additional elements and the transitions upon which these limits are based follow:  Na $<$-7.5 (Na I 5892.8, 5897.6); Si $<$-7.2 (Si I 3906.6); Sc $<$-9.2 (Sc II 3614.9): V $<$-8.5 (V II 3119.3); Cr $<$-8.2 (Cr I 3579.7, 3594.5, 3606.5; Cr II 3125.9, 3133.0); Mn $<$-8.1 (Mn II 3443.0).

\clearpage

\begin{table}
\caption\noindent{NLTT 43806 Absorption Lines, Element Abundances, and Mass Accretion Rates}
\begin{tabular}{@{}lcccc}
\br
Element& Wavelengths& log [Z/H] & log(t$_{set}$)& log(dM/dt)\\
& (\AA) & & (yr)& (g s$^{-1}$)\\
\mr
Na& 5891.6/5897.6& -8.1$\pm$0.14& 4.31& 6.95\\
Mg& 3830.4/3833.4/3839.4/5174.1/5185.0& -7.1$\pm$0.13& 4.28& 8.0\\
Al& 3945.1/3962.6& -7.6$\pm$0.17& 4.23& 7.6\\
Si& 3906.6& -7.2$\pm$0.14& 4.20& 8.04\\
Ca& 3934.8/3969.6/4227.9& -7.9$\pm$0.19& 4.14& 7.56\\
Ti& 3350.4/3362.2/3373.8/3384.7/3760.4& -9.55$\pm$0.14& 4.05& 6.08\\
Cr& 3594.5& -9.55$\pm$0.22& 4.02& 6.15\\
Fe& 3559.5/3566.4/3571.1/3582.2/3721.1/3723.6/3735.9& & &\\
& 3738.2/3746.6/3747.0/3750.6/3759/3/3764.8/3768.3&  &  &\\
& 3816.9/3821.5/3825.5/3827.0/3841.5/3851.0/3857.7& &  &\\
& 3861.0/4384.8& -7.8$\pm$0.17& 4.00& 7.95\\
Ni& 3381.7/3415.7/3459.4/3462.6/3494.0/3516.1/3525.5& -9.1$\pm$0.17& 3.97& 6.7\\
\br
\end{tabular}
\end{table}
\noindent Notes $-$ Wavelengths are in vacuum.  Abundance ratios are by number of atoms.  For all elements the listed (and for Fe also unlisted) detected transitions are from the neutral atoms with the exception of Ti for which all transitions are from Ti II and for Ca for which the 3934.8 and 3969.6 transitions are from Ca II.  Excepting Fe, all lines that appear in the HIRES spectra are listed.  For Fe we list only transitions that appear in Figures 2 to 4; an additional $\sim$20 transitions are detected, but are not listed.  The settling times and mass accretion rates (dM/dt) are as in the Note to Table 2.

\clearpage

\begin{table}
\caption\noindent{White Dwarfs With Eight or More Elements Heavier Than Helium Detected Optically}
\begin{tabular}{@{}lccccccc}
\br
Element& GD 362& J0738+1835& NLTT 43806& GD40& G241-6& PG1225-079& HS2253+8023\\
& DAZB& DBZA& DAZ& DBAZ& DBZ& DZAB& DBAZ\\
& 10,540 K& 13,600& 5,900 K& 15,300 K& 15,300 K& 10,800 K& 14,400 K\\
\mr
O& & x& & x& x& & x\\
Na& x& x& x& & & & \\
Mg& x& x& x& x& x& x& x\\
Al& x&  x& x& & & & \\
Si& x& x& x& x& x& & x\\
Ca& x& x& x& x& x& x& x\\
Sc& x& x& & & & x& \\
Ti& x& x& x& x&  x& x& x\\
V& x& x& & & & x& \\
Cr& x& x& x& x& x& x& x\\
Mn& x& x& & x& x& x& x\\
Fe& x& x& x& x& x& x& x\\
Co& x& ?& & & & & \\
Ni& x& x& x& & & x& \\
Cu& x& & & & & & \\
Sr& x& & & & & & \\
\br
\end{tabular}
\end{table}
\noindent Notes $-$ The white dwarf type and T$_{eff}$ are given under the name of each star:  GD 362 (Zuckerman et al 2007); J0738+1835 (P. Dufour, private communication 2011); NLTT 43806 (this paper); GD 40 (Klein et al 2010); G241-6 (Zuckerman et al 2010); PG1225-079 and HS2253+8023 (Klein 2011 and Klein et al 2011).  

\clearpage

\begin{landscape}
\begin{table}
\caption{White Dwarf Element Abundances Relative to Aluminum}
\begin{tabular}{@{}lcccccccccc}
\br
WD& Name& Type& T$_{eff}$& log g& log& Ca/Al& Mg/Al& Fe/Al& Si/Al& Ref.\\
& & & (K)& (cgs)&  [Ca/H(e)]& & & &\\
\br
solar& & & & & -5.66& 0.76& 12.3& 10.2& 12.0& 1\\
0208+396& G74-7& DAZ& 7,200& 7.93& -8.83& 1.42& 12.9& 12.9& & 2\\
0300-013& GD40& DBAZ& 15,300& 8.0& -6.88& $>$1.66& $>$7.2& $>$4.2& $>$2.2& 3\\
0322-019& G77-50& DAZ& 5,310& 8.05& -9.8& 0.25& 8.0& 3.2& $<$25& 4\\
0354+463& Rubin 80& DAZ& 7,800& 8.0& -8.33& 0.59& 11.8& 6.5& 19.4& 2\\
0419-487& LHS 1660& DAZ& 6,300& 8.54& -9.28& 0.31& 10.9& 5.6& 37.8& 2\\
0435+410& GD61& DBZA& 17,280& 8.2& -7.90& $>$0.25& $>$3.6& $>$0.28& $>$2.2& 5\\ 
J0738+1835& SDSSJ0738& DBZA& 13,600& 8.5& -6.8& 0.5& 63& 25& 40& 6\\
1257+278& G149-28& DAZ& 8,600& 8.1& -8.04& 1.35& 8.5& 5.75& $<$9.3& 7\\
1633+433& G180-63& DAZ& 6,600& 8.08& -8.63& 1.25& 15.0& 7.5& & 2\\
1653+385& NLTT 43806& DAZ& 5,900& 8.0& -7.9& 0.5& 3.16& 0.63& 2.5& 7\\
1729+371& GD362& DAZB& 10,500& 8.24& -6.24& 1.44& 2.63& 5.62& 3.63& 8\\
1929+012& GALEX1931& DAZ& 23,470& 7.99& -5.83& $>$1& $>$56& $>$56& $>$32& 9\\
2222+683& G241-6& DBZ& 15,300& 8.0& -7.25& $>$0.56& $>$5.1& $>$1.7& $>$1.66& 10\\
\br
\end{tabular}
\end{table}
\noindent Notes $-$  The 6th column ([Ca/H(e)] gives the log of the abundance ratio by number of Ca to the most abundant element, either H or He; for GD40, GD362 and SDSSJ0738 that element is He.  Columns 7-10 are ratios by number of atoms (and are not logs).  References for data are given in the right hand column: (1) Lodders 2003; (2) Zuckerman et al. 2003; (3) Klein et al. 2010; (4) Farihi et al. 2011b, P. Dufour 2011 (private comm.); (5) Farihi et al. 2011a; (6) Dufour et al. 2010, Dufour 2011 (private comm.); (7) the present paper; (8) Zuckerman et al. 2007: (9) Melis et al. 2011; (10) Zuckerman et al. 2010.  Vennes et al. (2010) give 20,890 K for T$_{eff}$ for WD1929+012 (see discussion of the temperature of this white dwarf in Melis et al. 2011). 
\end{landscape}

\clearpage

\begin{landscape}
\begin{table}
\caption{Element Abundances by Number in NLTT 43806, G149-28, and the Solar System}
\begin{tabular}{@{}lccccccccccc}
\br
Z/Al& NLTT& NLTT& G149-28&Solar& Bulk& Earth's& 30\% crust& lunar& CI& Euc& How\\
& 43806& 43806& photosphere& & Earth& continental& 70\% upper& mare& & & \\
& photosphere& steady state& & & & crust& mantle (by& basalt& & &\\
& & accretion& & & & & weight)& & & &\\
\mr
Ca/Al& 0.5& 0.61& 1.35& 0.76& 0.72& 0.425& 0.57& 1& 0.73& 0.76& 0.70\\
Mg/Al& 3.16& 2.82& 8.5& 12.3& 11.8& 0.428& 3.11& 0.77& 12.6& 0.74& 1.54\\
Fe/Al& 0.63& 1.07& 5.75& 10.2& 9.1& 0.405& 0.61& 1.31& 10.5& 1.10& 1.26\\
Si/Al& 2.5& 2.68& $<$9.3& 12.0& 11.0& 3.07& 4.48& 3.65& 12.0& 3.3& 4.1\\
Na/Al& 0.32& 0.27& $<$4.7& 0.69& 0.19& 0.32& 0.26& 0.043& 0.69& 0.038& 0.052\\
Ti/Al& 0.011& 0.0166& 0.05& 0.029& 0.0265& 0.036& 0.0343& 0.236& 0.029& 0.037& 0.031\\
Cr/Al& 0.011& 0.0178& $<$1& 0.155& 0.148& 0.001& 0.0147& & 0.158&  0.019& \\
Ni/Al& 0.032& 0.0575& 0.69& 0.57& 0.49& 0.00057& 0.0105& & 0.575&  & \\
\br
\end{tabular}
\end{table}
\noindent Notes $-$ The 30/70 Earth crust/mantle division is apportioned by weight.  However, all abundance ratio entries in the body of the table are by number of atoms, including those in the column headed "30\% crust 70\% upper mantle".  Abundance ratios for G149-28 in the steady state phase are approximately the same as the photospheric ratios because the various settling times are so similar (see Table 2).  See Sections 4.1 and 4.2 for explanation of the steady state accretion and crust/mantle columns.  Solar abundances from Lodders (2003);  bulk Earth from Allegre et al.(1995); Earth crust and upper mantle from Anderson (2007); CI meteorites from Lodders (2003) and Jarosewich (1990);  Eucrite and Howardite meteorites from Jarosewich (1990) and Kitts \& Lodders (1998).  The Ni abundance in Eucrites is quite variable (Kitts \& Lodders 1998).   See also Hawkesworth \& Kemp (2006) for slightly different average Earth crustal abundances.   
\end{landscape}

\clearpage

\begin{table}
\caption{Element Abundances in NLTT 43806 in a Declining Phase Model}
\begin{tabular}{@{}lcccc}
\br
Z/Al& NLTT 43806& Solar& Bulk Earth& GD 362\\
\mr
Na/Al& 0.16& 0.69& 0.19& 0.041\\
Mg/Al& 1.97& 12.3& 11.8& 2.63\\
Si/Al& 3.13& 12.0& 11.0& 3.63\\
Ca/Al& 1.25& 0.76& 0.72& 1.44\\
Ti/Al& 0.078& 0.029& 0.026& 0.029\\
Cr/Al& 0.125& 0.155& 0.15& 0.10\\
Fe/Al& 9.38& 10.2& 9.10& 5.62\\
Ni/Al& 0.78& 0.57& 0.49& 0.22\\
\br
\end{tabular}
\end{table}
\noindent Notes $-$  Abundance ratios are by number of atoms.  NLTT 43806 column gives photospheric abundance ratios 6.6 x 10$^4$ years ago (see Section 4.3).  Solar abundances from Lodders (2003);  bulk Earth from Allegre et al.(1995); GD 362 photospheric values from Zuckerman et al (2007).

\clearpage
\begin{figure}
\includegraphics[width=140mm]{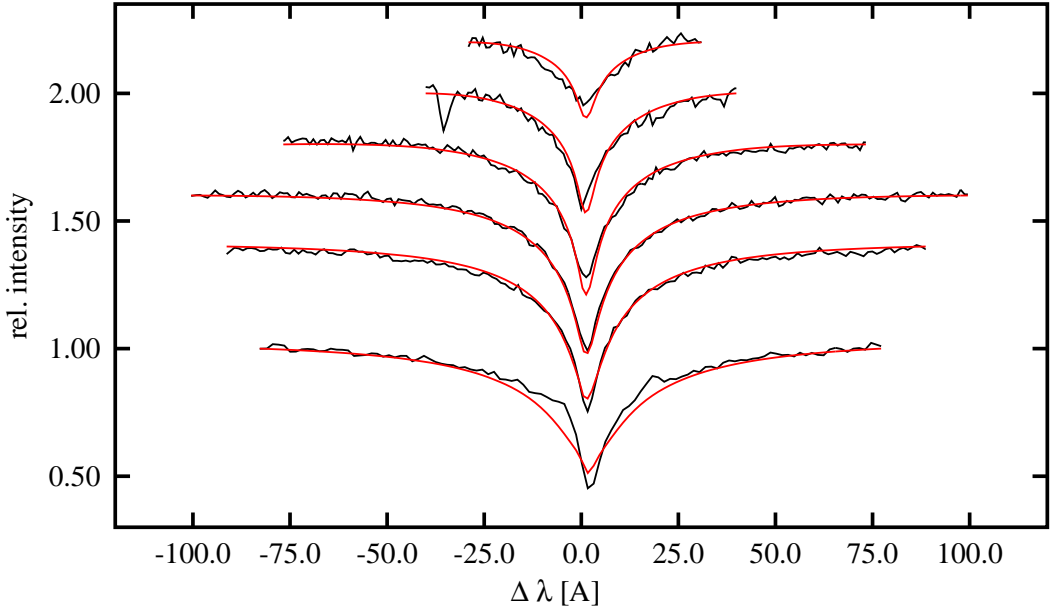}
\caption{\label{figure1} Fit to the Balmer lines H$\alpha$ to H8 in the SDSS spectrum of G149-28.}
\end{figure}

\clearpage
\begin{figure}
\includegraphics[width=140mm]{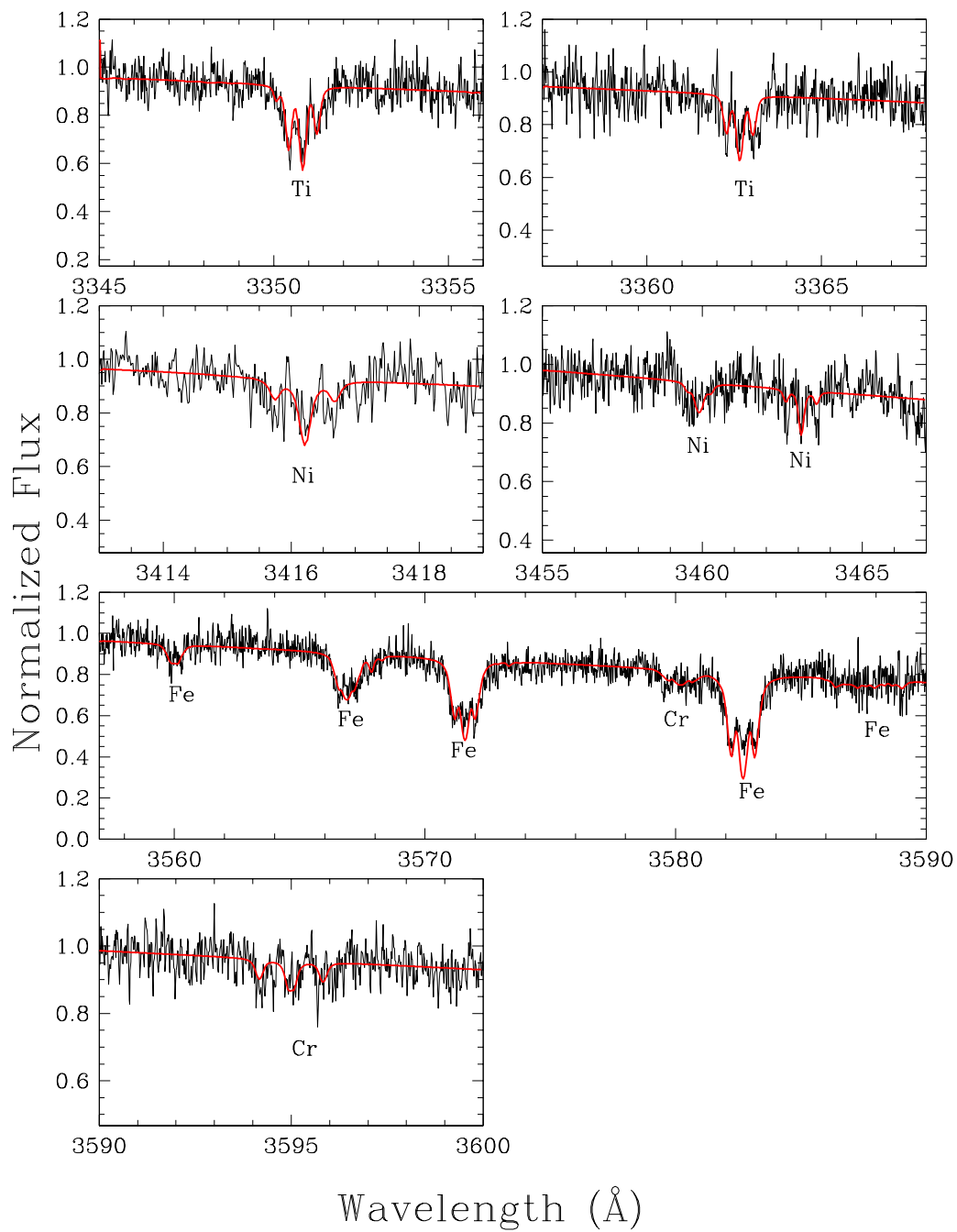}
\caption{\label{figure2} HIRES spectrum of NLTT 43806 overlaid with a model that includes a uniform 70 kG magnetic field.  The abscissa is wavelength in vacuum in the heliocentric rest frame.}
\end{figure}

\clearpage
\begin{figure}
\includegraphics[width=140mm]{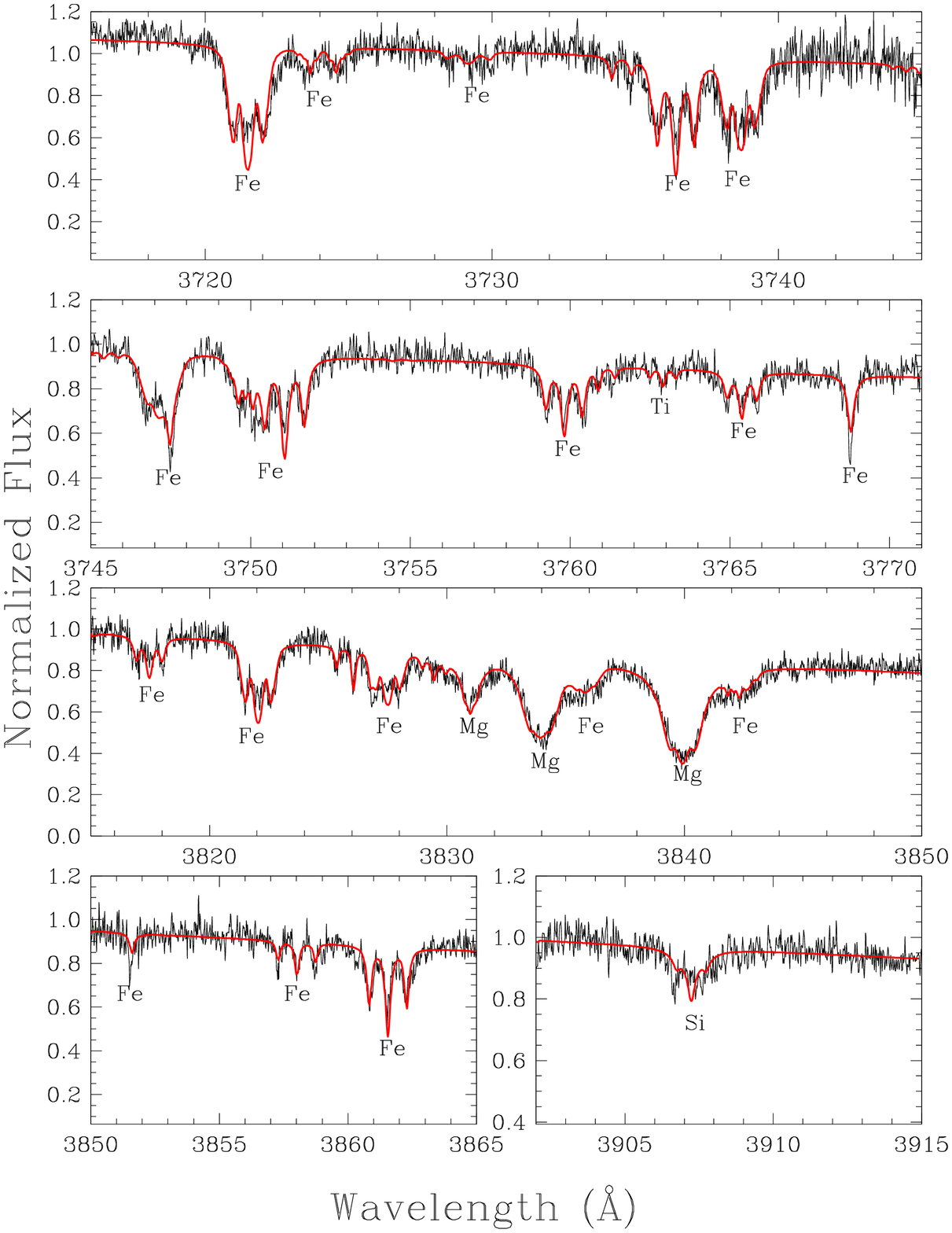}
\caption{\label{figure3} HIRES spectrum of NLTT 43806 overlaid with a model that includes a uniform 70 kG magnetic field.}
\end{figure}

\clearpage
\begin{figure}
\includegraphics[width=140mm]{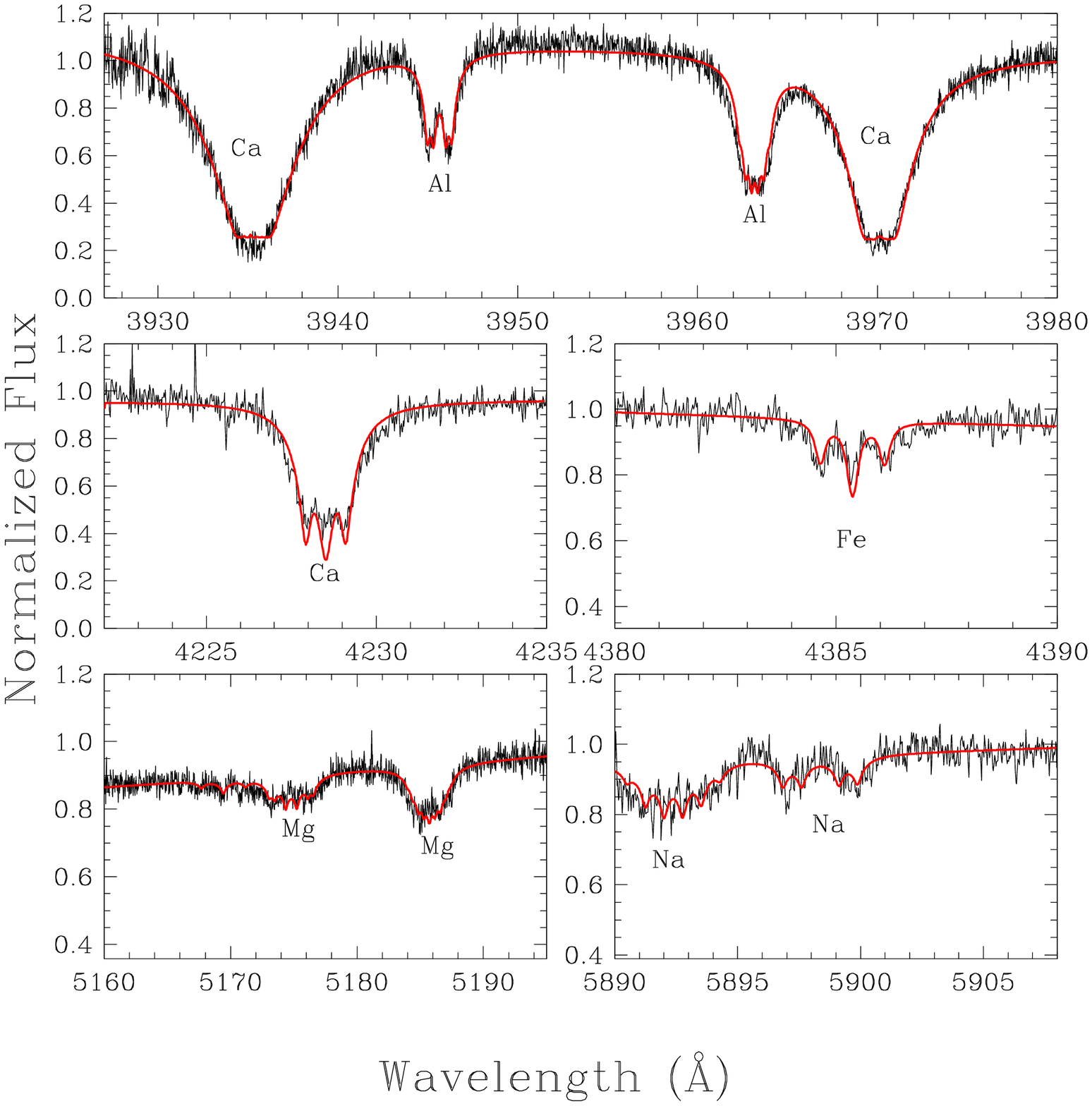}
\caption{\label{figure4} HIRES spectrum of NLTT 43806 overlaid with a model that includes a uniform 70 kG magnetic field.}
\end{figure}

\clearpage
\begin{figure}
\includegraphics[width=140mm]{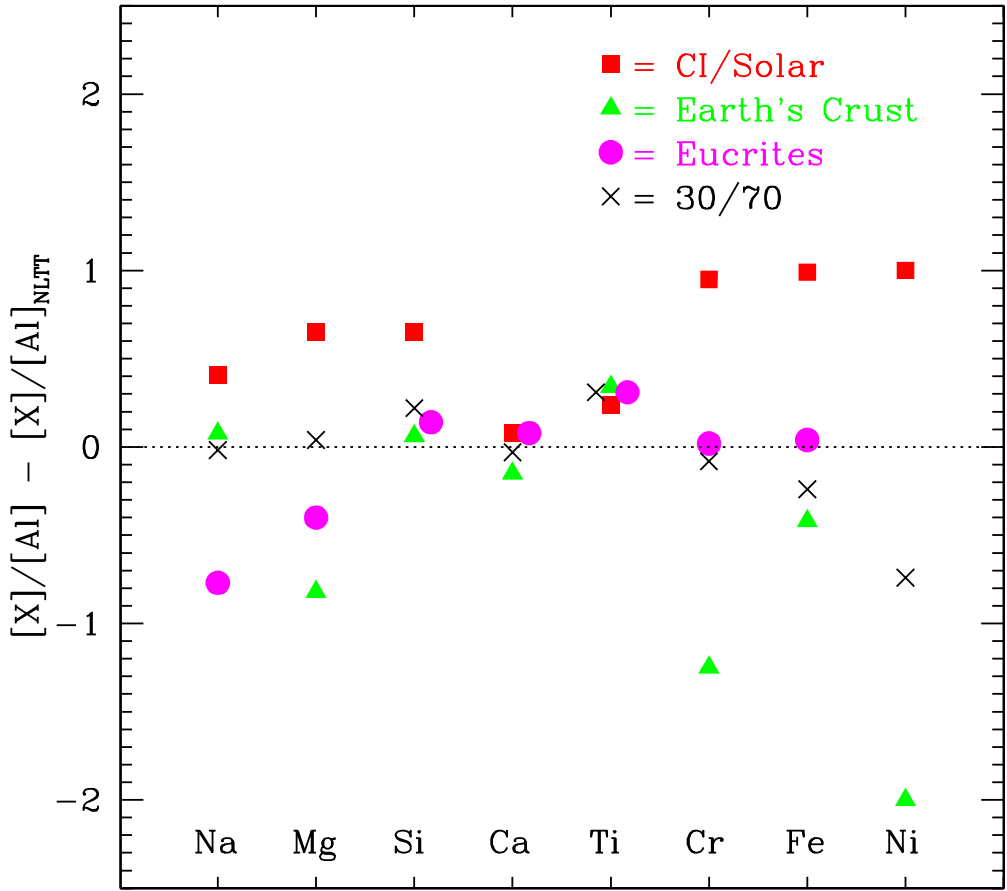}
\caption{\label{figure5} Element to Al abundance ratios for various models in Table 6 compared to the NLTT 43806 steady state accretion values.   The ordinate is the logarithm of an entry in Table 6 for a given model divided by the entry for the NLTT 43806 steady state model and where X signifies one of the 8 elements listed along the abscissa.   The 30/70 plotted points correspond to entries for the "30\% crust 70\% mantle" model.  The Eucrites plotted points are an average of the values listed in Table 6 under Euc and How.  As noted in Section 4.2, the models with the overall best agreement of element abundances with the NLTT 43806 data (that is, ordinate values near zero) are the 30/70 crust/mantle and the Euc/How mixtures.  Note that the bulk Earth abundances given in Table 6 would plot in Figure 5 at essentially the same positions as the CI/Solar entries, with the exception of Na that would plot at -0.15 rather than at +0.41.  Uncertainties in element abundance ratios in the NLTT 43806 steady state model (the denominator in the ratios plotted on the ordinate) are typically about 0.12 dex (see Section 3).   Variations in the element abundance ratios in CI and Eucrite meteorites and in Earth's crust and mantle (the numerator in the ratios plotted on the ordinate) may be evaluated from papers listed in the notes to Table 6 and papers cited therein.}   
\end{figure}

\end{document}